\documentclass[11pt,a4paper]{article}
\usepackage[T1]{fontenc}
\usepackage[margin=23mm]{geometry}
\usepackage{amsmath,amssymb,graphicx,booktabs,makecell}
\usepackage{newtxtext,newtxmath}
\usepackage{bm}
\usepackage{microtype}
\usepackage{adjustbox}
\usepackage{xcolor}
\usepackage{newunicodechar}
\usepackage[numbers,sort&compress]{natbib}
\usepackage{authblk}
\usepackage[hidelinks]{hyperref}

\numberwithin{equation}{section}
\newunicodechar{μ}{\ensuremath{\mu}}
\DeclareUnicodeCharacter{00B5}{\ensuremath{\upmu}}

\title{Bayesian reconstruction of quantum free energies from path-integral molecular dynamics}
\author[1,2]{Alexander V. Shapeev\thanks{Correspondence to: Alexander V. Shapeev, Skolkovo Institute of Science and Technology, Skolkovo Innovation Center, Bolshoy boulevard, 30, bld.~1, Moscow 121205, Russia. E-mail: \href{mailto:a.shapeev@skoltech.ru}{a.shapeev@skoltech.ru}. ORCID: \href{https://orcid.org/0000-0002-7497-5594}{0000-0002-7497-5594}.}}
\author[3]{Ivan S. Novikov}
\author[1,2]{Ekaterina Spirande}
\author[1,2]{Andrei Kolmakov}
\author[1,2,4]{Nikita Rybin}
\affil[1]{Skolkovo Institute of Science and Technology, Skolkovo Innovation Center, Bolshoy boulevard, 30, bld.~1, Moscow 121205, Russia}
\affil[2]{Digital Materials LLC, Kutuzovskaya, str.~4A, Odintsovo 143001, Russia}
\affil[3]{HSE University, Faculty of Computer Science, Pokrovsky boulevard 11, Moscow 109028, Russian Federation}
\affil[4]{National Research Nuclear University Moscow Engineering Physics Institute, Kashirskoe Highway 31, 115409, Moscow, Russia}
\date{}

\begin{document}
\maketitle

\begin{abstract}
Predicting the relative stability of molecular crystals requires free
energies that account for nuclear quantum motion and anharmonicity.
We present a Gaussian-process method that combines path-integral
molecular dynamics, phonon calculations, and physical constraints to
reconstruct Gibbs free-energy differences. The method enables
extrapolation to the quantum and thermodynamic limits and to zero
integration time step, quantifies uncertainty, and guides further
simulations through active learning. We demonstrate the method for
two benzene phases using a machine-learned interatomic potential.
The reconstructed free-energy advantage of the
orthorhombic phase decreases from $18.18\pm0.60$ meV per molecule
at zero temperature to $15.32\pm0.67$ meV per molecule at 200 K
($1\sigma$ uncertainties). Harmonic and
quantum-corrected classical approximations underestimate the reconstructed
gap, demonstrating the importance of treating anharmonicity and quantum
motion together, including their coupling.

\end{abstract}

\section{Introduction}\label{sec:1}

Nuclear quantum effects (NQE), such as zero-point motion and quantum
delocalization of nuclei, influence processes ranging from tunneling~\cite{PhysRevLett.125.216001} and thermal
conductivity~\cite{PhysRevLett.119.185901} to thermal expansion~\cite{kim2018nuclear}, phase behavior at extreme
conditions~\cite{PhysRevLett.110.065702}, and polymorph
stability~\cite{PhysRevLett.117.115702}. Molecular crystals are a prime
example: many molecules crystallize in several polymorphs,
whose relative stability at a given temperature and pressure is set by
their Gibbs free energies. Because the polymorph determines properties such
as solubility and processability, ranking polymorphs is a key goal of
crystal structure prediction, particularly for pharmaceutically-relevant
compounds~\cite{Price2014,Firaha2023}. Intramolecular modes, particularly those involving hydrogen, have vibrational energy larger than thermal energy ($\hbar\omega\gg k_BT$) and are dominated by zero-point motion, whereas intermolecular translations and librations are soft and anharmonic, with $\hbar\omega$ of the order of $k_BT$. The energy differences between polymorphs are frequently below 1~kcal/mol~\cite{CruzCabeza2015,Nyman2015}. Hence, performing computational ranking requires precise methods accounting for anharmonic nuclear motion and  NQE~\cite{Hoja2019,Firaha2023,Kapil2019anharmonic,PhysRevLett.117.115702,KapilEngel2022} -- path-integral molecular dynamics (PIMD) threats both on an equal footing.

Although PIMD is frequently used~\cite{markland2018nuclear,Ceriotti2016},
obtaining converged PIMD free energies of molecular crystals remains far
from routine. PIMD costs roughly $P$ times as much as classical molecular
dynamics, where $P$ is the number of ring-polymer beads per nucleus, and
free energies require long trajectories in large cells (supercells). Such
simulations therefore increasingly use machine-learned interatomic
potentials (MLIPs), which approach the accuracy of density-functional
theory (DFT) at a small fraction of its cost~\cite{Jacobs2025}. Even then,
the free energy is not an ensemble average and must be obtained by
thermodynamic integration, typically from a harmonic reference to the
anharmonic system and from classical to quantum
nuclei~\cite{Habershon2011,PhysRevLett.117.115702,KapilEngel2022}. Each
integration is performed at a fixed bead count, cell size, and time
step, so convergence in each parameter must be
established separately, with sampling noise complicating every
extrapolation~\cite{Liu2016,Korol2020}. The simulations must also be consistent with the reference
free energy, for example in the degrees of freedom they sample and the
time-step bias of the integrator, and the potential must be smooth enough
to reproduce the expected limiting behavior; violations of these
requirements are subtle and easily overlooked. Finally, because
free-energy differences between polymorphs are small, their statistical
uncertainty must be tracked and used to select further simulations.

An alternative to separate integrations and convergence studies is to treat
the free energy as an unknown function inferred from noisy simulation data,
as is done in Gaussian-process (GP) regression. This approach was used in
the previous works of some of us, where it has been shown how to combine
thermodynamic-integration, molecular dynamics, phonon, and melting-point
data into free energies and phase diagrams with quantified
uncertainties~\cite{Ladygin2021,Miryashkin2023,Klimanova2023,Spirande2025}. In the aforementioned works, however, the nuclear dynamics was treated classically, with quantum effects included at most at the harmonic or quasi-harmonic level~\cite{Spirande2025}. Extending to PIMD adds the bead count as a variable in which the low-temperature and quantum limits do not commute: at
finite bead count, the anharmonic free energy vanishes as $T\to0$, whereas
in the quantum limit it retains a zero-point contribution. In this work, we
develop this extension. We subtract an analytically known harmonic
reference, evaluated at the bead count and cell size of each simulation,
and relate the anharmonic remainder to the sampled enthalpies through an
exact thermodynamic identity. The quadratic vanishing of this remainder as
$T\to0$ at finite bead count and cell size supplies the integration
boundary condition, making the method a Bayesian, mesh-free thermodynamic
integration in temperature, and also tests the consistency of the
simulations with the reference. A joint GP model of the competing phases,
formulated in coordinates that resolve the non-commuting limits, combines
simulations at different temperatures, bead counts, cell sizes, and time
steps and extrapolates them jointly to the quantum and thermodynamic limits
and to zero time step, with correlated uncertainties for the free-energy
difference. Beyond selecting training data for the MLIP, active learning
then uses these uncertainties to choose the simulation conditions that most
reduce the uncertainty in the free-energy difference per unit computational
effort.

We apply the method to benzene -- a simple, prototypical molecular crystal
whose polymorphism has been studied extensively by experiment and
computation~\cite{Piermarini1969,Budzianowski2006,vanEijck1998,Azadi2016,KapilEngel2022}. At 1~bar and 0--200~K, we compare the experimental orthorhombic
phase~I~\cite{Budzianowski2006} with a tetragonal polymorph N5 from Ref.~\cite{vanEijck1998}, which was also predicted in
our previous crystal structure prediction study~\cite{Rybin2025}. Interatomic interactions are described by a moment tensor potential (MTP)~\cite{Shapeev2016,Novikov2021} fitted to DFT calculations. Step by step, the low-temperature condition reveals three inconsistencies, which we correct: degrees of freedom missing from the fixed-cell reference, the time-step bias of the integrator, and the non-smooth cutoff of the original MTP, removed by refitting it with smooth radial functions.

\section{Free-energy reconstruction method}\label{sec:2}

In this section, we present a Bayesian framework for reconstructing quantum
Gibbs free energies from path-integral molecular dynamics. We represent each
phase's free energy as an analytically known harmonic reference plus an
anharmonic contribution inferred from simulated enthalpies using a
Gaussian-process model. Requiring the anharmonic contribution to vanish
quadratically with temperature at finite bead count and cell size makes it
possible to determine free energies from enthalpy observations.
The model combines simulations across temperatures, bead counts $P$, and
system sizes $N$, automatically extrapolating to the quantum and
thermodynamic limits, $P,N\to\infty$, and quantifying uncertainty in the
free-energy difference between phases. This uncertainty guides further
simulations through active learning, targeting the greatest predicted
reduction in uncertainty per unit computational effort.

\subsection{Free energies and enthalpy observations}\label{sec:2.1}

Consider a crystalline phase $\alpha$ containing $N$ nuclei.
The cell volume and shape fluctuate
at fixed external pressure $p_{\mathrm{ext}}$. We denote temperature in kelvin
by $T$.
The nuclei interact through a total potential energy $U$
and are treated as distinguishable, neglecting nuclear exchange.

Our goal is to reconstruct the full Gibbs free energy per atom,
$G_\alpha^{\mathrm{full}}(T,P,N)$. Here $P$ is the number of
replicas, or beads, in the path-integral representation of the nuclei.
The case $P=1$ gives the classical free energy, while $P\to\infty$
recovers the quantum limit at fixed temperature. The quantity of interest
is the difference $\Delta G^{\mathrm{full}}(T)=G_{\mathrm{II}}^{\mathrm{full}}-G_{\mathrm I}^{\mathrm{full}}$
between two phases in the quantum and thermodynamic limits, $P,N\to\infty$.

For each phase, we construct a harmonic reference at a local minimum of
the static enthalpy $U+p_{\mathrm{ext}}V$ with respect to nuclear positions
and cell geometry, where $V$ is the cell volume. The reference contains
the static enthalpy at this minimum and the vibrational free energy,
evaluated analytically from the modes at the corresponding fixed cell
using the same bead count $P$ and particle number $N$ as the simulated
system. Overall translation and the quadratic
elastic contribution of the six homogeneous strain degrees of freedom
are included separately in the reference. Denoting its
free energy per atom by $G_\alpha^{\mathrm{ref}}$, we reconstruct
the anharmonic excess free energy
\begin{equation}
G_\alpha(T,P,N)
=G_\alpha^{\mathrm{full}}(T,P,N)-G_\alpha^{\mathrm{ref}}(T,P,N).
\end{equation}

Let $H_\alpha(T,P,N)$ denote the exact ensemble-mean enthalpy per atom.
Simulations estimate this quantity by averaging the sum of the bead-averaged
potential energy, the centroid-virial kinetic-energy estimator, and
$p_{\mathrm{ext}}V$, and dividing by $N$. At fixed pressure,
bead count, and particle number, the thermodynamic identity
$H_\alpha=G_\alpha^{\mathrm{full}}-T\partial_TG_\alpha^{\mathrm{full}}$
also applies to the reference. Subtracting the reference enthalpy
$H_{\mathrm{ref},\alpha}$ therefore gives
\begin{equation}
H_\alpha-H_{\mathrm{ref},\alpha}
=(1-T\partial_T)\bigl(G_\alpha^{\mathrm{full}}-G_\alpha^{\mathrm{ref}}\bigr)
=\mathcal L G_\alpha,
\end{equation}
where $\mathcal L:=1-T\partial_T$ is the enthalpy observation operator.
Temperature derivatives are taken at fixed pressure, bead count, and
particle number.

For a simulation condition $\ell$, specifying a phase $\alpha_\ell$,
temperature $T_\ell$, bead count $P_\ell$, and particle number $N_\ell$,
let $y_\ell$ denote the sampled mean enthalpy per atom minus the reference
enthalpy at the same condition. The observation used for reconstruction is
\begin{equation}
y_\ell=H_{\alpha_\ell}(T_\ell,P_\ell,N_\ell)
-H_{\mathrm{ref},\alpha_\ell}(T_\ell,P_\ell,N_\ell)+\epsilon_\ell
=(\mathcal L G_{\alpha_\ell})(T_\ell,P_\ell,N_\ell)+\epsilon_\ell,
\end{equation}
where $\epsilon_\ell$ is the sampling error in the mean enthalpy.
Adding a term proportional to $T$ to $G_\alpha$ leaves
$G_\alpha-T\partial_TG_\alpha$ unchanged. At fixed finite bead count and
system size, the low-temperature partition function is dominated by
fluctuations around a stable minimum. The reference includes the static
minimum enthalpy and the complete harmonic contribution. For a smooth
potential, the leading anharmonic correction to the free energy is
\begin{equation}
G_\alpha(T,P,N)=O(T^2),
\qquad T\to0,\quad P,N<\infty.
\end{equation}

This condition fixes the entropy contribution left undetermined by the
enthalpy observations.

\subsection{Coordinates for free-energy reconstruction}\label{sec:2.2}

If $P\to\infty$ is taken before $T\to0$, quantum fluctuations persist
and the excess free energy can retain a nonzero anharmonic zero-point
contribution. We seek coordinates that distinguish this limit from
cooling at finite $P$ and resolve the crossover between them.

\begin{figure}[tbp]
\centering
\includegraphics[width=1\linewidth,height=0.75\textheight,keepaspectratio]{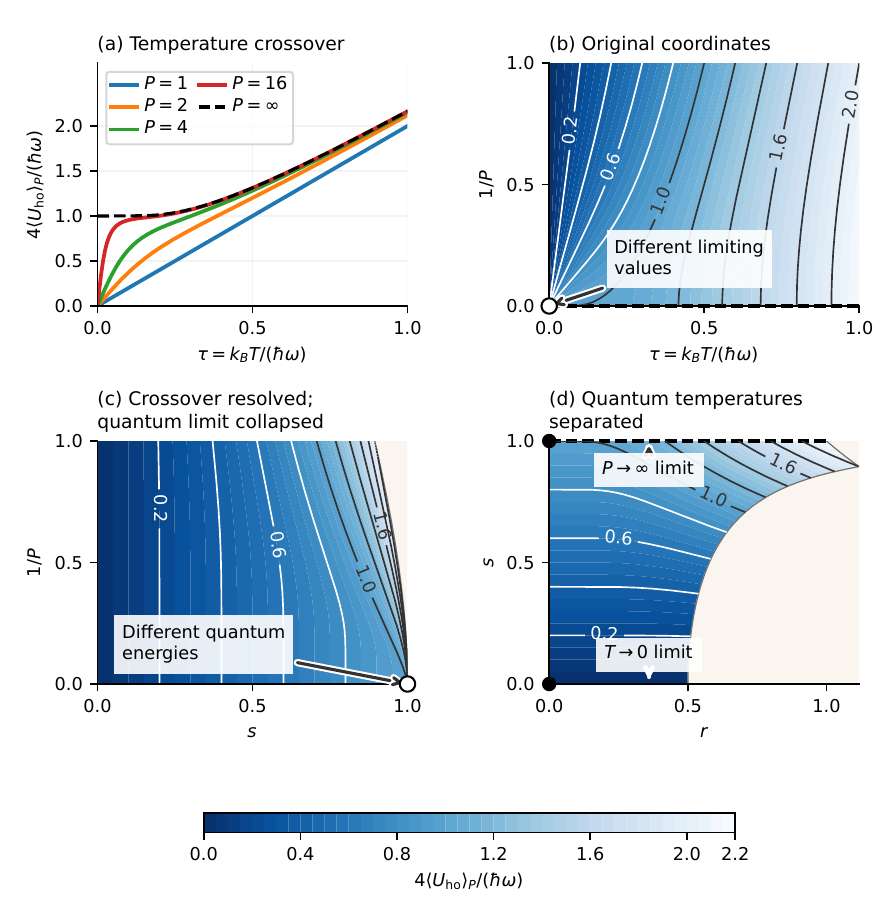}
\caption{Harmonic-oscillator motivation for the coordinates used to reconstruct
the excess free energy, with
$E_*=\hbar\omega$. (a) Normalized mean potential energy
$4\langle U_{\mathrm{ho}}\rangle_P/(\hbar\omega)$ at finite bead counts and
in the quantum limit. (b–d) Contours of the same quantity in
$(\tau,1/P)$, $(s,1/P)$, and $(r,s)$, respectively. Open circles mark
points without a unique limiting energy. Filled circles in (d) mark
$(r,s)=(0,0)$ and $(0,1)$, with normalized energies zero and one, obtained
by cooling before and after taking $P\to\infty$, respectively. Dashed
black curves show the quantum limit. The contour panels share the same
levels, color scale, and physical domain, $0\le\tau\le1$, $P\ge1$;
beige regions lie outside this domain. Contours use the analytic
oscillator expression continued to real $P\ge1$.}
\label{fig:kernel_coordinate_construction}
\end{figure}

Consider a harmonic oscillator of angular frequency $\omega$, with energy
quantum $\hbar\omega$. Figure~\ref{fig:kernel_coordinate_construction}a shows its normalized mean
potential energy $4\langle U_{\mathrm{ho}}\rangle_P/(\hbar\omega)$ as a function
of the dimensionless temperature $\tau=k_BT/(\hbar\omega)$ for different
bead counts $P$, where $k_B$ is Boltzmann's constant.
Here $\langle U_{\mathrm{ho}}\rangle_P$ denotes the exact ensemble average
of the bead-averaged potential energy in the finite-$P$ harmonic model.
At finite $P$, this energy vanishes as $\tau\to0$,
whereas its quantum limit tends to one. The crossover occurs within a
temperature interval of width $\tau=O(P^{-1})$. This shrinking boundary
layer appears as contours converging toward the origin in $(\tau,1/P)$
(Figure~\ref{fig:kernel_coordinate_construction}b).

Taking the normalized boundary-layer profile as a coordinate gives
\begin{equation}
s=\frac{\tau}{\sqrt{\tau^2+1/(4P^2)}}.
\end{equation}

In $(s,1/P)$, the low-temperature contours remain separated as $P$
increases (Figure~\ref{fig:kernel_coordinate_construction}c). However, quantum limits at different positive
temperatures bring different energies to the same point,
$(s,1/P)=(1,0)$. To preserve this temperature dependence, we pair $s$
with the radial coordinate
\begin{equation}
r=\sqrt{\tau^2+\frac{1}{4P^2}},
\qquad \tau=rs.
\end{equation}

The $1/P^2$ term makes both coordinates approach their quantum limits
with $O(P^{-2})$ corrections at fixed positive $T$. A smooth model in
these coordinates can therefore reproduce the leading bead-discretization
error of the primitive path-integral approximation.

The quantum limit now lies on $s=1$, with $r=\tau$, while finite-$P$
cooling paths end at $(r,s)=(1/(2P),0)$ (Figure~\ref{fig:kernel_coordinate_construction}d). Cooling before or
after taking $P\to\infty$ therefore leads to $(r,s)=(0,0)$ or $(0,1)$,
respectively.

Guided by this harmonic example, we use the same coordinates for the
anharmonic excess free energy of a crystal, replacing $\tau$ by
$\theta=k_BT/E_*$, where $E_*>0$ is a common energy scale for both phases,
fitted together with the other model hyperparameters (Section~\ref{sec:2.3}). The condition
$G_\alpha=O(T^2)$ becomes $G_\alpha=O(s^2)$ at $s=0$, leaving the quantum
zero-point value free.

For cell size, we use $\xi=N_{\mathrm{base},\alpha}/N$, where
$N_{\mathrm{base},\alpha}$ is the number of atoms in the base simulation
cell of phase $\alpha$. The thermodynamic limit is $\xi=0$. We model the
excess free energies as smooth functions of $\mathbf x=(r,s,\xi)$.

In the full reconstruction, the integration time step $\Delta t$ is an
additional coordinate, and the target free energy is obtained in the
joint limit $P,N\to\infty$, $\Delta t\to0$. For the benzene phase
difference studied here, we find no detectable residual time-step
dependence after correcting the harmonic integration bias (Section~\ref{sec:5.4}).
We therefore use the reduced model in $\mathbf x=(r,s,\xi)$ for the
principal reconstruction. Section~\ref{sec:2.3} gives the covariance including
the time step, and Section~\ref{sec:5.4} tests this simplification at the present precision.

\subsection{Gaussian-process reconstruction of free energies}\label{sec:2.3}

We assign a joint zero-mean Gaussian-process (GP) prior to the two excess
free energies, with a shared contribution and independent phase-specific
contributions. For inputs $\mathbf x=(r,s,\xi)$ and
$\tilde{\mathbf x}=(\tilde r,\tilde s,\tilde\xi)$, their covariance before
imposing the quantum zero-temperature constraint below is
\begin{equation}
\operatorname{Cov}[G_\alpha(\mathbf x),G_{\alpha'}(\tilde{\mathbf x})]
=k_S(\mathbf x,\tilde{\mathbf x})+\delta_{\alpha\alpha'}k_D(\mathbf x,\tilde{\mathbf x}),
\end{equation}
where $k_S$ and $k_D$ are the shared and phase-specific kernels, and
$\delta_{\alpha\alpha'}$ is the Kronecker delta. We construct the kernels
from squared-exponential covariance factors, expressing smooth variation
with inverse cell size $1/N$ and the coordinates $r$ and $s$. Writing the
squared-exponential kernel as
$k_\ell(u,\tilde u)=\exp[-(u-\tilde u)^2/(2\ell^2)]$, we choose, for $a\in\{S,D\}$,
\begin{equation}
\begin{aligned}
k_a(\mathbf x,\tilde{\mathbf x})
&=k_{\ell_\xi}(\xi,\tilde\xi)\left[
A_a^2 k_{\ell_r}(r,\tilde r)Q_a(s,\tilde s)+B_a^2s^2\tilde s^2
\right],\\
Q_a(s,\tilde s)
&=s\tilde s\left[k_{\ell_{s,a}}(s,\tilde s)
-k_{\ell_{s,a}}(s,0)k_{\ell_{s,a}}(0,\tilde s)\right].
\end{aligned}
\end{equation}

The covariance $Q_a$ describes a smooth process conditioned to vanish at
$s=0$ and multiplied by $s$, enforcing the required $O(s^2)$ behavior.
The $B_a$ term adds a contribution
proportional to $s^2$ that is independent of $r$. The amplitudes $A_a,B_a$ and correlation lengths $\ell_{s,a}$ are
component-specific; $\ell_r$ and $\ell_\xi$ are shared. These four amplitudes
and four correlation lengths, together with $E_*$, are the nine
hyperparameters controlling the reduced GP prior.

We further condition the joint process on
$\partial_rG_\alpha(0,1,\xi)=0$ for both phases and every $\xi$, leaving
the zero-point values free. This expresses vanishing
excess entropy at zero temperature in the bulk quantum limit.
The cell in our simulations is shared by all beads and is not path-integrated.
Extending the same zero-slope condition to finite sizes assumes that reference
subtraction leaves no residual zero-temperature excess entropy from cell
fluctuations.

Denote the covariance obtained after imposing these constraints by
$K_{\alpha\alpha'}(\mathbf x,\tilde{\mathbf x})$. To include the
integration time step, we extend this covariance to
\begin{equation}
K^{(\Delta t)}_{\alpha\alpha'}
\big((\mathbf x,\Delta t),(\tilde{\mathbf x},\Delta t')\big)
=K_{\alpha\alpha'}(\mathbf x,\tilde{\mathbf x})
\left[1+(\kappa\Delta t)^2(\kappa\Delta t')^2\right].
\end{equation}

This adds a contribution quadratic in $\Delta t$ while preserving the
physical constraints. The additional hyperparameter $\kappa\ge0$ has units of inverse
time; the added contribution's prior amplitude relative to the
time-step-independent part is $(\kappa\Delta t)^2$. Setting $\kappa=0$
recovers the reduced model. With the full covariance, extrapolation to
zero time step is obtained by evaluating the posterior at $\Delta t=0$.

We apply the enthalpy observation operator $\mathcal L$ to this covariance and
use independent Gaussian sampling errors with standard deviations
$\sigma_\ell=1.5\,\mathrm{SE}_\ell$, where $\mathrm{SE}_\ell$ is the standard
error of the sampled mean enthalpy, estimated from ten (occasionally more)
independent trajectory averages. The factor 1.5 accounts approximately for uncertainty in the
estimated standard error. For ten independent, normally distributed
trajectory averages, it gives a Gaussian two-sided 99.9\% confidence interval
slightly wider than the corresponding Student-$t$ interval with nine degrees
of freedom.
Section~\ref{sec:5.2} compares this prescription with an explicit Student-$t$
observation model.
We report Gaussian $\pm2\sigma$ intervals as 95\% intervals, with Student-$t$
intervals evaluated at the same probability level.

The kernel hyperparameters and $E_*$ are fitted by maximizing the marginal likelihood,
with the Gaussian prior $\log\ell_\xi\sim\mathcal N(\log1.5,1)$ added
to facilitate active learning.

The quantity of interest is the phase free-energy difference in the
quantum and thermodynamic limits,
\begin{equation}
\Delta G^{\mathrm{full}}(T)
=\lim_{P,N\to\infty}
\big[G_{\mathrm{II}}^{\mathrm{full}}(T,P,N)
-G_{\mathrm I}^{\mathrm{full}}(T,P,N)\big].
\end{equation}

The chosen coordinates allow the fitted GP model to take both limits
analytically: simultaneous extrapolation from finite bead counts and cell
sizes reduces to evaluating the posterior mean and covariance at
$\mathbf x_T=(k_BT/E_*,1,0)$. This gives the limiting excess free energies
and their uncertainties. Adding the harmonic references in the same limits
yields $\Delta G^{\mathrm{full}}(T)$.

We employ active learning to select further simulations for free-energy
reconstruction, keeping the interatomic potential fixed. This procedure is
separate from the active learning used to select training configurations
for the potential (Section~\ref{sec:3.3}). We rank candidates by their
predicted reduction in the variance of $\Delta G^{\mathrm{full}}$ per unit
effort, targeting the temperature with the largest current variance.
For a candidate $j$ at
$(T_j,P_j,N_j)$, we model the sampling variance and additional effort as
\begin{equation}
\widehat\sigma_j^2=\frac{aT_j^2}{N_j n_j},
\qquad
\mathcal E_j=N_jP_j(n_j+n_{j,\mathrm{equil}}),
\end{equation}
respectively, where $n_j$ and $n_{j,\mathrm{equil}}$ count additional
production and equilibration steps, summed over trajectories.
The common coefficient $a$ is the median of
$\sigma_\ell^2N_\ell n_\ell/T_\ell^2$ over observed
conditions in both phases, using their accumulated production steps
$n_\ell$.

\section{Benzene computational setup}\label{sec:3}

We apply the free-energy reconstruction method to two crystalline packings
of benzene, $\mathrm{C}_6\mathrm{H}_6$, and compare their Gibbs free energies.
Both structures are described by the same machine-learned interatomic
potential, fitted to density-functional theory (DFT) calculations. We
sample each packing with path-integral molecular dynamics at different
temperatures, bead counts, and cell sizes to determine how quantum and
anharmonic nuclear motion affect their relative stability.
We used the Atomica software package to set up and run the atomistic
calculations.

In the benzene application, crystal energies, enthalpies, and free energies
are reported per molecule unless stated otherwise. We retain the symbols
of Section~\ref{sec:2}; their per-molecule values are twelve times the per-atom values.
The particle number $N$ continues to count atoms.

Further computational details are given in the Supplementary Information.

\subsection{Crystal structures}\label{sec:3.1}

\begin{figure}[tbp]
\centering
\includegraphics[width=1\linewidth,height=0.75\textheight,keepaspectratio]{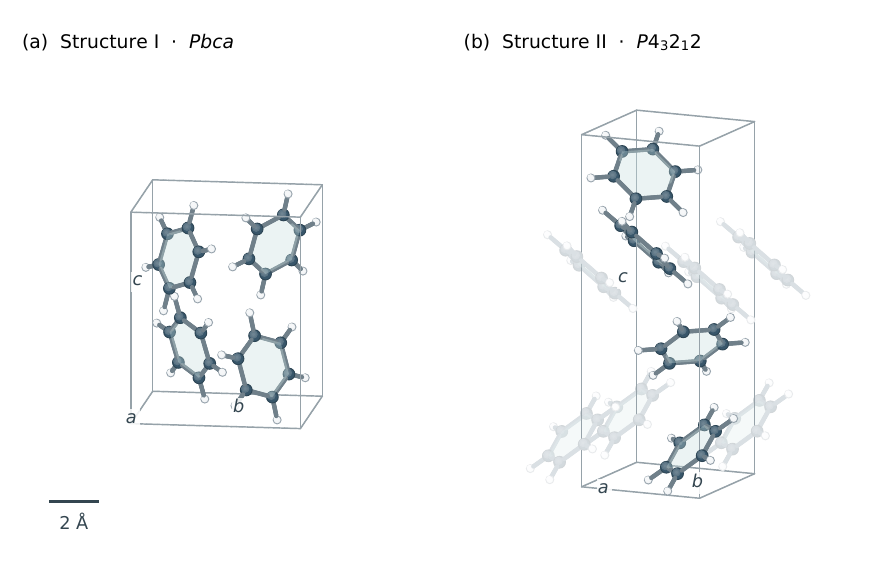}
\caption{Starting configurations of (a) the orthorhombic $Pbca$ packing
and (b) the tetragonal $P4_32_12$ packing, before relaxation with the fitted
potential. Carbon atoms are blue-grey and hydrogen atoms are white;
shading within the carbon rings highlights their planes. Each cell contains
four molecular centres, whose complete molecules are drawn in full colour.
Additional periodic copies with carbon atoms inside the displayed cell are
shown faintly. Cell edges and lattice axes are indicated. Both panels use
orthographic projection and the same length scale; atom sizes are chosen
for visibility. The displayed cell origins are shifted to reduce the
number of molecules crossing cell boundaries.}
\label{fig:benzene_crystals}
\end{figure}

Our starting configurations come from our earlier crystal structure
prediction study~\cite{Rybin2025}. We denote the orthorhombic $Pbca$ polymorph by structure I and the tetragonal $P4_32_12$ polymorph by structure II. The phase indices I and II in the free-energy reconstruction refer to the crystalline basins associated with these two polymorphs, respectively. These labels do not denote the experimental phase sequence. Each crystallographic cell contains four molecules, or 48 atoms. Structure I corresponds to the experimental benzene I reported by
Budzianowski and Katrusiak~\cite{Budzianowski2006},
with inclined molecular rings forming a herringbone arrangement.
Structure II corresponds to the predicted structure N5 of
van Eijck et al.~\cite{vanEijck1998}. Its fourfold screw
symmetry relates molecules by rotation and translation along the
elongated cell axis (see Figure~\ref{fig:benzene_crystals}).

\subsection{Density-functional calculations}\label{sec:3.2}

We generated the DFT training data and static structural references with the
Vienna Ab initio Simulation Package (VASP)~\cite{Kresse1996a},
using the Perdew–Burke–Ernzerhof (PBE) exchange-correlation functional~\cite{Perdew1996} and the
projector augmented-wave (PAW) method~\cite{Blochl1994,Kresse1999}.
Dispersion interactions were included through the DFT-D3 correction with
zero damping~\cite{Grimme2010}.

The plane-wave energy cutoff was 550 eV. Brillouin-zone sampling used
$\Gamma$-centered meshes with a reciprocal-space spacing parameter of
0.2 Å$^{-1}$. We used Gaussian smearing of the electronic occupations
with a width of 0.05 eV.

\subsection{Moment tensor potentials}\label{sec:3.3}

We generated the training database with a level-16 moment tensor potential
(MTP)~\cite{Shapeev2016,Novikov2021}. The level controls
the complexity of the many-body basis. We trained the potential on DFT
energies, forces, and stresses for both crystal packings. Configurations
were selected by active learning during relaxations of crystals with
perturbed atomic positions and cell vectors, followed by classical
molecular dynamics in the NPT ensemble at 1 bar and temperatures of
100 and 200 K. The resulting database contained
650 configurations. We then fitted MTPs of levels 8, 12, and 16 to this
common database, using the same 6 Å cutoff and radial basis.

To assess their suitability for the free-energy comparison, we examined
the static energy difference between the relaxed structures (II minus I),
alongside training errors and evaluation cost (Table~\ref{tab:1}). Level 8 gave
only 2.0 meV per molecule, compared with 22.7 meV per molecule from DFT.
Levels 12 and 16 gave 24.5 and 21.9 meV per molecule, respectively.
Level 16 reduced the training errors further but required about 1.7 times
the evaluation cost of level 12. We selected level 12 as the original
potential; the calculations at 1 bar use its smooth refit (Section~\ref{sec:5}).

\begin{table}[tbp]
\centering
\caption{Training errors, evaluation times, and relaxed structures for the three
MTPs fitted to the same 650-configuration database. MAE and RMSE denote
mean absolute error and root-mean-square error. Evaluation times cover
energies, forces, and virials on one CPU core. All MTPs use a 6 Å cutoff
and 12 Chebyshev radial functions. The static energy difference $\Delta u$
is the energy of structure II relative to structure I after relaxation.
For each MTP and structure family, the lowest-energy configuration found
in the relaxation search is reported.
$\ast$ For MTP-12, slight changes in molecular orientation lower the symmetry
from $P4_32_12$ to $P4_3$, while the cell retains the tetragonal geometry
compatible with $P4_32_12$.}
\label{tab:1}
\small
\begin{adjustbox}{max width=\linewidth}
\begin{tabular}{lrrrr}
\toprule
Quantity & MTP-8 & MTP-12 & MTP-16 & DFT (VASP) \\
\midrule
Parameters & 107 & 175 & 286 & — \\
Evaluation time (µs/atom) & 8.4 & 14.1 & 24.2 & — \\
Energy MAE (meV/molecule) & 23.7 & 9.3 & 7.0 & — \\
Energy RMSE (meV/molecule) & 35.6 & 13.2 & 9.5 & — \\
Force RMSE (meV/Å) & 135.1 & 63.2 & 42.3 & — \\
Stress RMSE (GPa) & 0.60 & 0.23 & 0.15 & — \\
I: space group & $\bm{Pbca}$ & $\bm{Pbca}$ & $\bm{Pbca}$ & $\bm{Pbca}$ \\
II: space group & $\bm{P4_32_12}$ & $P4_3^{\ast}$ & $P4_3$ & $\bm{P4_32_12}$ \\
$\Delta u$ (meV/molecule) & 2.0 & \textbf{24.5} & \textbf{21.9} & \textbf{22.7} \\
\bottomrule
\end{tabular}
\end{adjustbox}
\end{table}

\subsection{Path-integral simulations}\label{sec:3.4}

We performed path-integral molecular dynamics using an interface between i-PI~\cite{Kapil2019} and MLIP-2~\cite{Solovykh2025}. We sampled the isothermal–isobaric ensemble with the PILE-G thermostat
and a flexible-cell barostat, allowing both the volume and shape of the
cell to fluctuate. We use 16 MPa for the correction tests and 1 bar for the
final free-energy calculations (Sections~\ref{sec:4} and~\ref{sec:5}).
The standard time step was 0.5 fs. We tested its
effect on the corrected enthalpies and reconstructed phase free-energy difference
using continuations at 0.25 and 0.125 fs (Section~\ref{sec:5.4}).

We screen the final configurations of the PIMD trajectories to exclude
those showing signs of a phase change, as phase II appears close to melting
at 200 K.

\section{Validation and refinement}\label{sec:4}

Applying the method in practice required several refinements. We illustrate
these by reconstructing the free energies of the two benzene polymorphs
from PIMD trajectories generated at 16 MPa.\footnote{The pressure was inadvertently set to $10^{-4}$ eV Å$^{-3}$
    (approximately 16 MPa) instead of the intended $10^{-4}$ GPa (1 bar).}

\begin{figure}[tbp]
\centering
\includegraphics[width=1\linewidth,height=0.75\textheight,keepaspectratio]{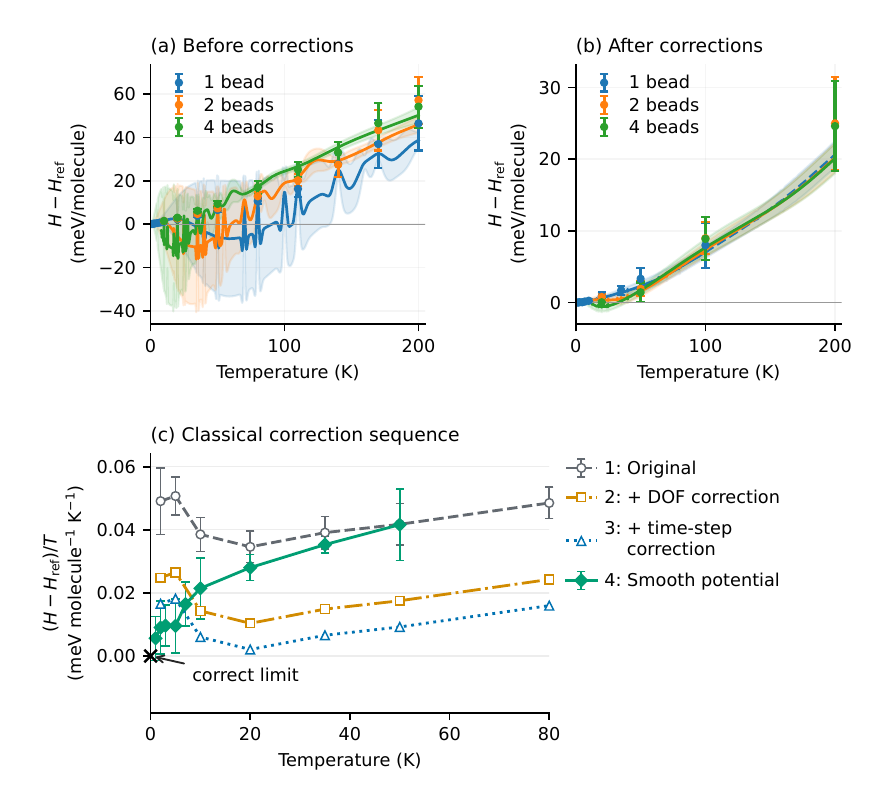}
\caption{Excess enthalpies $H-H_{\mathrm{ref}}$ at 16 MPa and a 0.5 fs time step.
(a,b) Phase I with four molecules: (a) original potential with direct harmonic
subtraction; (b) smooth potential
after degrees-of-freedom and time-step corrections.
Both joint fits use observations from
both phases at all available cell sizes and bead counts. Points have error
bars $\pm2\sigma_\ell$; curves and bands show posterior means and $\pm2$
posterior standard deviations conditional on fitted hyperparameters. Dashed
segments extend beyond the sampled temperature range.
(c) Classical ($P=1$) excess enthalpy divided by temperature for phase II
with sixteen molecules. Points show sampled estimates and lines join them.
Curves 1--3 use the same trajectories with successive degrees-of-freedom and
time-step corrections;
curve 4 uses the smooth potential, its own trajectories
and harmonic reference, and both corrections. Error bars
$\pm2\sigma_\ell/T$ are shown for curves 1 and 4; omitted bars for curves 2
and 3 have the same size as those for curve 1. The marked origin,
labelled ``correct limit'', shows the required $T\to0$ limit of the ratio.}
\label{fig:before_after_phase_I}
\end{figure}

Initially, we subtracted only the fixed-cell harmonic reference from the
sampled enthalpies and fitted the joint free-energy model to both phases.
The resulting fit is unsatisfactory, with large uncertainty and unphysical oscillations between observations (Figure~\ref{fig:before_after_phase_I}a).
We therefore examined the corrections needed to make the
sampled enthalpies consistent with the reference defined in Section~\ref{sec:2}.

A closer look at the low-temperature region of Figure~\ref{fig:before_after_phase_I}a reveals the
underlying problem: the excess enthalpy data do not follow the expected
$O(T^2)$ behavior at fixed bead count. This is clearer in Figure~\ref{fig:before_after_phase_I}c,
which shows the classical excess enthalpy divided by temperature for
phase II. The ratio should vanish as $T\to0$, but the original data
(dashed curve) instead level off well above zero.

This behavior indicates a deviation from the classical harmonic enthalpy
contribution of $3k_BT$ per atom. A natural place to look
is the degrees of freedom sampled in the simulations but
absent from the fixed-cell harmonic reference. Overall translation and
the six homogeneous cell strains contribute $9k_BT/2$ per simulation cell.
The \emph{degrees-of-freedom correction} subtracts this contribution from the
sampled enthalpy, shifting the ratio downward
(Figure~\ref{fig:before_after_phase_I}c, second curve), but does not fully remove the low-temperature offset.

Comparisons at different time steps provided the next clue.
A finite integration time step biases the sampled enthalpy even for a
harmonic system~\cite{Liu2016}; in the classical case, the leading bias is
$O((\Delta t)^2T)$. We calculated this contribution for each vibrational
mode of the reference crystal following Korol et al.~\cite{Korol2020} and subtracted the sum from
the sampled enthalpy. This \emph{time-step correction} further reduces the low-temperature offset
(Figure~\ref{fig:before_after_phase_I}c, third curve), but does not eliminate it.

After removing the harmonic time-step bias, the expected $O(T^2)$ remainder
relies on a smooth expansion of the potential beyond its quadratic part.
However, the original MTP radial basis has discontinuous second derivatives
at the cutoff.
Refitting with smooth radial functions produces a \emph{smooth potential}.
The resulting low-temperature trend in the excess enthalpy divided by
temperature is consistent,
within the observation uncertainties, with a line through the origin
(Figure~\ref{fig:before_after_phase_I}c, fourth curve). With all corrections applied, the joint fit
no longer shows unphysical oscillations, and its uncertainty
bands are substantially narrower (Figure~\ref{fig:before_after_phase_I}b). We therefore use this smooth
potential for the calculations at 1 bar described in Section~\ref{sec:5}.
Explicit formulas for the harmonic reference and enthalpy corrections are
collected in the Supplementary Information.

\section{Free-energy reconstruction for benzene}\label{sec:5}

We now apply the method to the two benzene phases at 1 bar, using
the smooth potential and corrections established above, to determine how
anharmonicity and nuclear quantum motion affect their relative stability.
All energies below are reported per molecule.

To construct the smooth potential, we retained the level-12 MTP and
replaced its radial functions by functions whose derivatives of every
order vanish at the cutoff. We increased the radial basis from 12 to 20
functions and the cutoff from 6 to 6.5 Å. After initializing the new
functions by approximating the original ones, we refitted all 271
parameters to the same 650-configuration DFT database. The resulting
potential reduces the training errors and recovers the $P4_32_12$ symmetry
of structure II, with a static energy difference of 23.6 meV per molecule,
compared with 22.7 meV per molecule from DFT.
Table~\ref{tab:2} compares the fitted potentials.

\begin{table}[tbp]
\centering
\caption{Training errors, evaluation times, and relaxed structures for the
original and smooth potentials. All errors are evaluated on the same
650-configuration DFT database. Timings are per energy, force, and virial
evaluation on one CPU core. The original-potential and DFT results are
repeated from Table~\ref{tab:1}. The DFT energy difference is obtained from fresh
single-point calculations on the relaxed geometries. The smooth-potential
structural values refer to the relaxed reference configurations used for sampling.
The energy differences contain no
vibrational contributions.}
\label{tab:2}
\small
\begin{adjustbox}{max width=\linewidth}
\begin{tabular}{lrrrr}
\toprule
Quantity & MTP-12 & MTP-16 & Smooth MTP-12 & DFT (VASP) \\
\midrule
Cutoff (Å) & 6.0 & 6.0 & 6.5 & — \\
Radial functions & 12 & 12 & 20 & — \\
Radial-function smoothness & $C^1$ & $C^1$ & $C^\infty$ & — \\
Parameters & 175 & 286 & 271 & — \\
Evaluation time (µs/atom) & 14.1 & 24.2 & 28.7 & — \\
Energy MAE (meV/molecule) & 9.3 & 7.0 & 8.2 & — \\
Energy RMSE (meV/molecule) & 13.2 & 9.5 & 11.0 & — \\
Force RMSE (meV/Å) & 63.2 & 42.3 & 53.3 & — \\
Stress RMSE (GPa) & 0.23 & 0.15 & 0.16 & — \\
I: space group & $\bm{Pbca}$ & $\bm{Pbca}$ & $\bm{Pbca}$ & $\bm{Pbca}$ \\
II: space group & $P4_3$ & $P4_3$ & $\bm{P4_32_12}$ & $\bm{P4_32_12}$ \\
$\Delta u$ (meV/molecule) & \textbf{24.5} & \textbf{21.9} & \textbf{23.6} & \textbf{22.7} \\
\bottomrule
\end{tabular}
\end{adjustbox}
\end{table}

\subsection{Free-energy reconstruction and active learning}\label{sec:5.1}

The initial dataset for active learning comprised 40 conditions at 1 bar:
22 for phase I and 18 for phase II. In the smallest cells, containing
4 and 16 molecules respectively, we sampled $T=10,25,50,100,150,200$ K
with $P=1,2,4$, except that phase II at 200 K included only $P=1$.
The initial dataset also included classical ($P=1$) simulations at 10 and
100 K in larger cells: 32 and 108 molecules for phase I and
128 molecules for phase II. The sampling protocol is the same as in
Section~\ref{sec:3.4}.

For each structure, we relaxed the atomic positions and cell at 1 bar and
calculated new harmonic vibrational spectra. We subtracted the fixed-cell
harmonic enthalpy and applied the degrees-of-freedom and time-step
corrections to the sampled enthalpies. We fitted the joint free-energy model
of Section~\ref{sec:2}.

We then used the active learning procedure described at the end of
Section~\ref{sec:2.3} to select additional simulation conditions and
extend existing trajectories. The final dataset contains 69 combinations of
phase, temperature, bead count, and cell size, with 720 independent
trajectories in total. Temperatures range from 1.25 to 200 K, with bead
counts $P=1,2,4,8$. The simulation cells contain 4, 32, or 108 molecules
for phase I and 16, 128, or 432 for phase II.

\begin{figure}[tbp]
\centering
\includegraphics[width=1\linewidth,height=0.75\textheight,keepaspectratio]{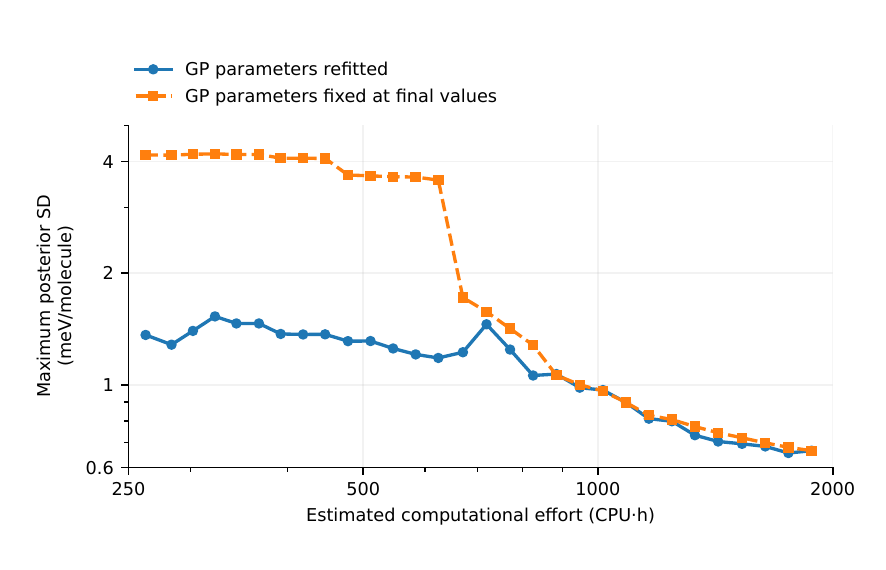}
\caption{Uncertainty in the reconstructed free-energy difference during
active learning. Each point gives the maximum posterior standard
deviation of the quantum, infinite-system phase free-energy difference on
a 250-point temperature grid spanning 5–200 K. Model hyperparameters are either
refitted at each selected point or held at their final values. Trajectories
are truncated to the data available at each selected point, with the final
trajectory screening applied throughout. The cumulative effort includes
equilibration and subsequently discarded trajectories. CPU·h estimates
use the single-core potential-evaluation timing in Table~\ref{tab:2} (28.7 μs per
atom per bead per step), excluding integration, communication, and model
fitting overheads.}
\label{fig:sigma_vs_cpu_hours}
\end{figure}

We reconstructed the free energies at 30 selected points during the
active-learning procedure, finding an overall reduction in the maximum
uncertainty of the quantum, infinite-system phase free-energy difference
as data accumulate (Figure~\ref{fig:sigma_vs_cpu_hours}).
With the model hyperparameters refitted at each selected point, the maximum
posterior standard deviation falls from 1.36 to 0.67 meV per molecule.
Repeating the reconstruction with the final model hyperparameters held
fixed separates the effect of additional data from changes in the fitted
hyperparameters.

\subsection{Quantum free-energy difference}\label{sec:5.2}

We extrapolated both excess free energies to $P,N\to\infty$. Adding the
harmonic references extrapolated to these limits gives the full quantum
Gibbs free-energy difference, denoted here by
$\Delta G_{\mathrm q}^{\mathrm{full}}=G_{\mathrm{II,q}}^{\mathrm{full}}-G_{\mathrm{I,q}}^{\mathrm{full}}$.

The reconstruction favors phase I throughout 0–200 K (Figure~\ref{fig:comparison_with_student_t}, green curve).
Its free-energy advantage decreases from $18.18\pm0.60$ meV per molecule
at zero temperature to $15.32\pm0.67$ meV per molecule at 200 K.
The quoted uncertainties are one posterior standard deviation, including
the covariance between the two phase free energies. They describe the
reconstruction with the fitted model hyperparameters and harmonic
references held fixed and do not include errors in the interatomic potential.

Replacing the Gaussian error model, with standard errors multiplied by 1.5,
by a Student's $t$ model
(Figure~\ref{fig:comparison_with_student_t}, black dashed curve)
narrows the 95\% posterior intervals by 12\% on average over 0–200 K,
while leaving the median prediction nearly unchanged.

\subsection{Quantum motion and anharmonicity}\label{sec:5.3}

\begin{figure}[tbp]
\centering
\includegraphics[width=1\linewidth,height=0.75\textheight,keepaspectratio]{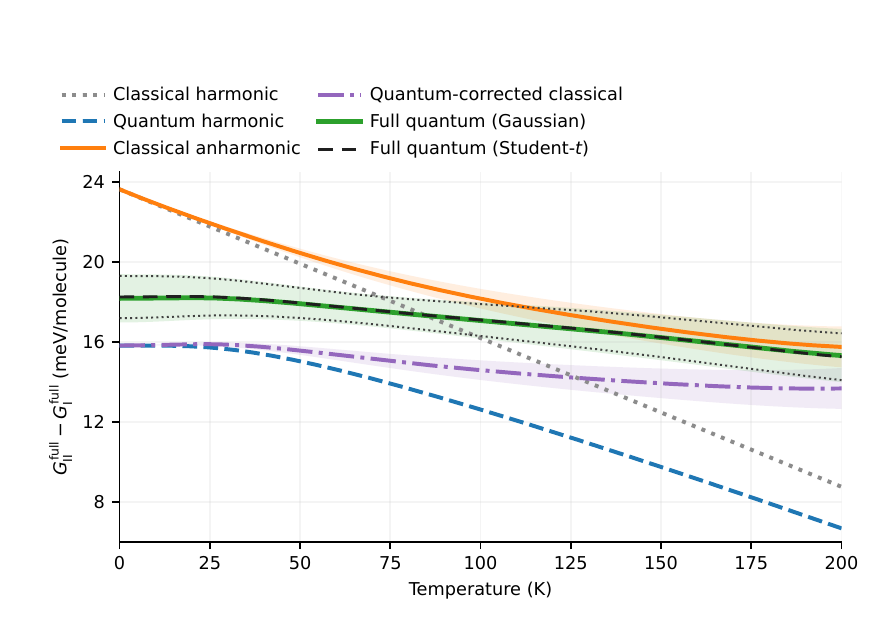}
\caption{Phase free-energy difference at 1 bar in the thermodynamic limit,
comparing physical approximations and sampling-error models. The
quantum-corrected classical curve combines classical anharmonicity with
the quantum harmonic reference. The full quantum result is shown as the
Gaussian posterior mean (green) and Student-$t$ posterior median (black
dashed). Shading denotes ±2 Gaussian posterior standard deviations;
black dotted curves bound the Student-$t$ 95\% intervals. Positive
values favor phase I. Model hyperparameters and harmonic references are held
fixed in the sampling-error comparison.}
\label{fig:comparison_with_student_t}
\end{figure}

Figure~\ref{fig:comparison_with_student_t} compares the full quantum result with harmonic and classical
calculations for the same potential. The quantum harmonic approximation
underestimates the gap by 2.4 meV per molecule at zero temperature and
8.6 meV per molecule at 200 K. At the upper end of this range, it captures
less than half of the gap. Anharmonicity thus substantially strengthens
the preference for phase I and reduces its temperature dependence.

Classical simulations include anharmonic motion but omit nuclear quantum
effects. We also consider an approximation that replaces the classical
harmonic contribution by its quantum counterpart, giving
\begin{equation}
\Delta G_{\mathrm{qc}}^{\mathrm{full}}
=\Delta G_{\mathrm{cl}}^{\mathrm{full}}
-\Delta G_{\mathrm{cl}}^{\mathrm{harm}}
+\Delta G_{\mathrm q}^{\mathrm{harm}},
\end{equation}
where the subscripts cl and q denote classical and quantum nuclei, the
superscript harm indicates the harmonic approximation, and qc denotes the corrected result.
This combines classical anharmonicity with harmonic quantum effects~\cite{Spirande2025}.
We obtain the classical anharmonic contribution from the
$P=1$, $N\to\infty$ limit of the same joint reconstruction.

The correction improves on the harmonic approximation at finite
temperature, but still underestimates the quantum gap by about 1.7–2.5 meV
per molecule over 0–200 K. The remaining difference is the contribution
from coupling between quantum motion and anharmonicity. At 200 K, the
uncorrected classical result is closer to PIMD because the harmonic
quantum correction and this coupling largely cancel. Thus adding only
the harmonic quantum correction can worsen the agreement; treating
quantum effects and anharmonicity together matters even when the classical
free-energy difference happens to be close to the quantum result.

None of the approximations considered consistently reproduces the
reconstructed quantum free-energy difference over the studied temperature
range. This comparison demonstrates the importance of treating quantum
motion and anharmonicity together, including their coupling.

\subsection{Time-step convergence}\label{sec:5.4}

The harmonic correction introduced in Section~\ref{sec:4} removes the integration
bias of the harmonic reference, but a residual bias may remain in the
anharmonic simulations. We tested this by reducing the time
step from 0.5 fs to 0.25 and 0.125 fs at selected conditions for both
structures. After applying the harmonic correction at each time step,
the excess enthalpies agree within their sampling uncertainties.

To determine how any remaining time-step dependence affects the
free-energy difference, we combined the 23 additional observations
with the 69 observations of Section~\ref{sec:5.1} and fitted the GP using the
time-step covariance of Section~\ref{sec:2.3}.

The best fit favors no time-step dependence (time-step sensitivity
$\kappa\approx0$), but nonzero dependence remains compatible with the
data. We therefore
averaged the predictions over possible values of $\kappa$ to include
this uncertainty in the reconstructed free-energy difference.

The additional observations also refine the phase-difference estimate.
Using all 92 observations and extrapolating the time step to zero gives
a zero-temperature gap of $17.57\pm0.53$ meV per molecule, compared with
$18.18\pm0.60$ in Section~\ref{sec:5.2}.
This shift mainly reflects the additional sampling. To quantify the
contribution from time-step extrapolation, we compare predictions at zero
and 0.5 fs from the same fit to all 92 observations:
\begin{equation}
D(T)=\Delta G_{\mathrm q}^{\mathrm{full}}(T;\Delta t=0)
-\Delta G_{\mathrm q}^{\mathrm{full}}(T;\Delta t=0.5\,\mathrm{fs}).
\end{equation}

We obtain $D(0)=-0.03\pm0.17$ meV per molecule.
Thus, no residual time-step dependence is detected over 0–200 K at the present
precision.

Fitting details, sampling-error sensitivity, and time-step convergence
results are given in the Supplementary Information.

\section{Discussion and conclusions}\label{sec:6}

We have developed a method for Bayesian reconstruction of polymorph
free-energy differences from PIMD observations and physical constraints.
The harmonic contribution is known analytically at each bead count,
leaving the anharmonic contribution to be determined from the simulations.
The low-temperature behavior constrains this unknown remainder: at any
fixed finite bead count, the anharmonic free energy vanishes quadratically
as the temperature approaches zero.
The method can thus be viewed as a Bayesian, mesh-free form of thermodynamic
integration in temperature: we infer the anharmonic free energy from
enthalpy observations, with its low-temperature behavior supplying the
integration boundary condition. The harmonic reference supplies the
analytically known contribution at each temperature.

The full method includes cell size, bead count, and integration time step
as coordinates, allowing extrapolation to $P,N\to\infty$ and
$\Delta t\to0$. Given an interatomic potential, the method is free
of systematic bias in the large-data limit with converged equilibrium
sampling. The reconstruction uses an exact thermodynamic
relation and does not truncate anharmonic interactions or their coupling
to quantum motion; extrapolation to zero time step also removes the
integration bias. In the present application, the explicit time-step
study detects no residual dependence after the harmonic correction
(Section~\ref{sec:5.4}), supporting the reconstruction without time-step
extrapolation used for the principal results.

For finite datasets, however, the achievable accuracy depends on the
quantity of interest. In the present calculation, active learning selected
bead numbers no larger than $P=8$, yielding an estimated uncertainty of
0.6–0.7 meV per molecule in the polymorph free-energy difference. The
absolute quantum free energies of the individual phases may still require
substantially larger bead numbers to resolve anharmonic zero-point
contributions from high-frequency intramolecular modes, such as C–H
stretching. These contributions are expected to produce similar shifts
in the two polymorphs, largely cancelling in their difference even while
the individual absolute values remain unconverged.

For benzene with the interatomic potential used here, the reconstruction
favors phase I throughout 0–200 K, with its free-energy advantage
decreasing from about 18.2 to 15.3 meV per molecule. The quantum harmonic
approximation increasingly underestimates this gap as temperature rises.
Adding classical anharmonicity improves the agreement, but a harmonic
quantum correction to the classical free energy still misses the coupling
between quantum motion and anharmonicity. This corrected classical
approximation performs better at low temperature, while the uncorrected
classical result is closer at high temperature because omitted
contributions largely cancel. None of these approximations consistently
reproduces the PIMD free-energy difference over the studied temperature
range.

Extending the method to other molecular crystals would allow quantum and
anharmonic contributions to be assessed in materials with competing
polymorphs. Potential applications include ranking pharmaceutical
polymorphs, studying how polymorphism affects solubility, and assessing
the thermodynamic stability of organic crystals. Combining the
reconstruction of polymorph free-energy differences with solid–liquid
coexistence calculations,
such as the melting-point
approach of Klimanova et al.~\cite{Klimanova2023},
could enable the construction of phase diagrams encompassing both
solid–solid and solid–liquid transitions. In these applications,
uncertainty estimates and active learning could reduce the computational
effort needed to resolve the relative stability of closely competing
phases.

\section*{Declarations}

\subsection*{Authors\textquotesingle{} contributions}

Conceptualization and methodology: N.R., A.V.S.

Software development: A.V.S., E.S., A.K.

Computational simulations: A.V.S., I.S.N.

Formal analysis and validation: A.V.S., N.R.

Visualization: A.V.S.

Writing -- original draft: A.V.S., I.S.N.

Writing -- review and editing: all the authors

Supervision and funding acquisition: A.V.S.

\subsection*{Availability of data and materials}

The data supporting this study are deposited at
\url{https://gitlab.com/ashapeev/pimd2-data}.
The repository contains crystal structures, fitted interatomic potentials
and their training data, harmonic references, scalar trajectory records
and endpoint configurations, screening results, processed observations,
free-energy predictions, and active-learning histories and computational
costs. File formats, units, and dataset coverage are documented in the
repository.

\subsection*{Declaration of Generative AI Use}

We made extensive use of AI tools in conducting this research
and preparing the manuscript.

OpenAI's Codex (Sol 6 and Astra 6) and Anthropic's Claude Code (Opus 5 and Fable 5) were used interchangeably
for code development and automation of computational workflows.
These tools were used to adapt the Gaussian-process code of
Spirande et al.~\cite{Spirande2025},
schedule PIMD calculations on our workstations, and collect the results.
Decisions about which data to collect were made by the authors or by our
active-learning algorithms, rather than by generative AI tools.

Codex was also used to generate natural-language reports from computational
output, including text and figures. The authors edited these reports for
accuracy, reviewed and verified their content, and used them as a basis
for writing the manuscript. The figures were generated using Python scripts from
simulation results, crystal structures, and analytical calculations.
AI tools assisted with the scripts; no image-generation models were used.

Codex also assisted with text polishing and with drafting and checking the
derivation of the coordinates for free-energy reconstruction (Section~\ref{sec:2.2}).

The authors reviewed and verified the AI-assisted computational
results and mathematical derivations, and take full responsibility
for the scientific content and conclusions of the manuscript.

\subsection*{Financial support and sponsorship}

This work was supported by the Russian Science Foundation (grant number 23-13-00332-P,
\url{https://rscf.ru/project/23-13-00332/}).
The funder had no role in the study design, data collection and analysis, interpretation of results, manuscript preparation, or decision to submit the manuscript for publication.

\subsection*{Conflicts of interest}

All authors declare that there are no conflicts of interest.

\bibliography{references}

\end{document}

\begin{center}
{\large\bfseries Supplementary Information}\\[4pt]
{\large Bayesian reconstruction of quantum free energies\\
from path-integral molecular dynamics}
\end{center}

This supplement documents the implementation and validation of the
free-energy reconstruction method. Section~\ref{sec:reference} collects
explicit formulas for implementing the harmonic references, enthalpy
corrections, and Gaussian-process (GP) prediction and active-learning
procedures introduced in the main text. Section~\ref{sec:protocol} documents
the training database for the interatomic potential, the benzene
calculations at 1 bar, sensitivity to the sampling-error model, and
time-step convergence.

\section{Reference quantities and free-energy reconstruction formulas}
\label{sec:reference}

All crystal energies and free energies in this supplement are per molecule;
$N$ counts atoms and $M=N/12$ counts molecules. The phase label is
$\alpha\in\{\mathrm I,\mathrm{II}\}$, $P$ is the bead count, $T$ is the
temperature, and $\beta=(k_BT)^{-1}$. The nuclei are distinguishable,
and sampling is restricted to the chosen crystalline basin.
We denote the full Gibbs free energy and ensemble-mean enthalpy by
$G_\alpha^{\mathrm{full}}(T,P,N)$ and $H_\alpha(T,P,N)$, respectively,
with pressure held fixed.

\paragraph{Harmonic reference.}
We write $p=p_{\mathrm{ext}}$ for the imposed external pressure.
For each phase, let $(\mathbf R_p,\mathbf h_p)$ be the relaxed atomic
positions and cell at this pressure, and let
$U_p$ and $V_p$ be the corresponding whole-cell potential energy and
volume. The positive vibrational frequencies at this fixed cell are
$\omega_\lambda$, $\lambda=1,\ldots,3N-3$, with the three translations
removed. Define $z_\lambda=\beta\hbar\omega_\lambda/(2P)$. The harmonic
Gibbs free energy and enthalpy at finite $P$, including the static
contribution, are
\begin{align}
G_\alpha^{\mathrm{harm}}
&=\frac{U_p+pV_p}{M}
+\frac{k_BT}{M}\sum_{\lambda=1}^{3N-3}
\log\!\left[2\sinh\!\left(P\operatorname{arsinh}z_\lambda\right)\right],
\label{eq:harmonic-free}\\
H_{\mathrm{harm},\alpha}
&=\frac{U_p+pV_p}{M}
+\frac{1}{M}\sum_{\lambda=1}^{3N-3}
\frac{\hbar\omega_\lambda\coth(P\operatorname{arsinh}z_\lambda)}
{2\sqrt{1+z_\lambda^2}}.
\label{eq:harmonic-enthalpy}
\end{align}
The enthalpy observation operator $\mathcal L=1-T\partial_T$, with
the temperature derivative taken at fixed $P,N,p$, relates the two
expressions by $H_{\mathrm{harm},\alpha}=\mathcal L G_\alpha^{\mathrm{harm}}$.
The classical and quantum references follow
by taking $P=1$ and $P\to\infty$, respectively.
Adding overall translation and the six quadratic homogeneous strain
coordinates gives the complete reference, with Gibbs free energy
$G_\alpha^{\mathrm{ref}}$ and enthalpy $H_{\mathrm{ref},\alpha}$.
These modes contribute another $9k_BT/2$ to the reference enthalpy per cell. Thus
$H_{\mathrm{ref},\alpha}=H_{\mathrm{harm},\alpha}+9k_BT/(2M)$.
The strain term is elastic enthalpy; fictitious barostat kinetic energy
does not enter the measured enthalpy. The complete reference uses the same
cell measure and translational convention as the sampled system. Its
global-mode free energy vanishes per molecule in the thermodynamic limit.
The GP infers the anharmonic excess Gibbs free energy
$G_\alpha=G_\alpha^{\mathrm{full}}-G_\alpha^{\mathrm{ref}}$.

\paragraph{Corrected observations.}
At condition $\ell$, let $\widehat h_\ell$ be the sampled mean of
$(\overline U+K_{\mathrm{CV}}+pV)/M_\ell$,
where $\overline U$ is the bead-averaged potential energy and
$K_{\mathrm{CV}}$ is the whole-cell centroid-virial kinetic-energy estimator:
\begin{equation}
K_{\mathrm{CV}}=\frac{3Nk_BT}{2}
-\frac{1}{2P}\sum_{i=1}^{N}\sum_{k=1}^{P}
(\mathbf R_i^{(k)}-\mathbf R_i^c)\cdot\mathbf f_i^{(k)},
\qquad \mathbf R_i^c=\frac{1}{P}\sum_{k=1}^{P}\mathbf R_i^{(k)}.
\end{equation}
Here $\mathbf f_i^{(k)}=-\nabla_iU(\mathbf R^{(k)})$ is the physical force.
The observation supplied to the GP is
\begin{equation}
y_\ell=\widehat h_\ell
-H_{\mathrm{harm},\alpha_\ell}(T_\ell,P_\ell,N_\ell)
-\frac{9k_BT_\ell}{2M_\ell}
-\frac{1}{M_\ell}\sum_{\lambda=1}^{3N_\ell-3}
\delta E_{\mathrm{CV}}(\omega_\lambda;T_\ell,P_\ell,\Delta t_\ell).
\label{eq:observation}
\end{equation}
The last term corrects the harmonic integration bias~\cite{Liu2016,Korol2020}.
The remaining
observation is modeled as $y_\ell=(\mathcal L G_{\alpha_\ell})_\ell+
\epsilon_\ell$, where $\epsilon_\ell$ is sampling error.
Residual time-step dependence is tested in Section~\ref{sec:protocol}.

\paragraph{Harmonic integration-bias correction.}
The i-PI calculations use OBABO splitting: half thermostat (O) and physical
force (B) updates, full free-ring-polymer propagation (A), then half B and O
updates. With physical bead masses and exact A propagation, we use the
harmonic stationary distribution of OBABO~\cite{Liu2016,Korol2020}
to obtain the following correction.
For a vibrational mode of the reference crystal with angular frequency
$\omega$, define the free-ring-polymer frequencies $\nu_j$, associated
with the bead springs, and the total harmonic frequencies $\Omega_j$,
which also include the crystal restoring force:
\begin{equation}
\nu_j=\frac{2P}{\beta\hbar}\sin\frac{\pi j}{P},
\qquad \Omega_j^2=\omega^2+\nu_j^2,
\qquad j=0,\ldots,P-1.
\end{equation}
The effective frequencies $w_j$ describing the stationary coordinate
distribution at integration time step $\Delta t$ are
\begin{equation}
w_j^2=\nu_j^2+\omega^2(\nu_j\Delta t)\cot(\nu_j\Delta t)
-\frac{\omega^4(\Delta t)^2}{4}\quad(j\ge1),
\qquad w_0^2=\omega^2\left[1-\frac{\omega^2(\Delta t)^2}{4}\right].
\end{equation}
The mode $j=0$ is the centroid mode. The mass-weighted ring-polymer
normal coordinates have stationary variances
$P/(\beta w_j^2)$, compared with $P/(\beta\Omega_j^2)$ at zero time step.
Their contribution to the centroid-virial total-energy bias is
\begin{equation}
\delta E_{\mathrm{CV}}(\omega;T,P,\Delta t)
=\frac{1}{2\beta}\left(\frac{\omega^2}{w_0^2}-1\right)
+\frac{1}{\beta}\sum_{j=1}^{P-1}
\left(\frac{\omega^2}{w_j^2}-\frac{\omega^2}{\Omega_j^2}\right).
\label{eq:time-step}
\end{equation}
This formula assumes stable, nonresonant steps with $w_j^2>0$ and
equilibration to the stationary distribution. It is exact for the fixed-cell
harmonic model; its use for anharmonic trajectories neglects anharmonic
changes in the bias and coupling to cell dynamics. It retains the finite-$P$
discretization effects that the GP subsequently extrapolates.

\paragraph{Conditioned covariance.}
Use $\mathbf x=(r,s,\xi)$, where
\begin{equation}
\theta=\frac{k_BT}{E_*},\qquad
r=\sqrt{\theta^2+\frac{1}{4P^2}},\qquad
s=\frac{\theta}{r},\qquad \xi=\frac{N_{\mathrm{base},\alpha}}{N}.
\end{equation}
Here $E_*>0$ is fitted, and $N_{\mathrm{base},\mathrm I}=48$ and
$N_{\mathrm{base},\mathrm{II}}=192$ are the base simulation-cell sizes.
Writing $k_\ell(u,u')=\exp[-(u-u')^2/(2\ell^2)]$, define the shared
and phase-specific kernels, indexed by $a\in\{S,D\}$, as
\begin{align}
Q_a(s,s')&=ss'\left[k_{\ell_{s,a}}(s,s')
-k_{\ell_{s,a}}(s,0)k_{\ell_{s,a}}(0,s')\right],\\
k_a(\mathbf x,\mathbf x')&=k_{\ell_\xi}(\xi,\xi')
\left[A_a^2k_{\ell_r}(r,r')Q_a(s,s')+B_a^2s^2s'^2\right].
\end{align}
These kernels enforce $G_\alpha=O(s^2)$, or $O(T^2)$ at finite $P,N$.
The unconditioned phase covariance is $k_S+\delta_{\alpha\alpha'}k_D$.
To impose $\partial_rG_\alpha(0,1,\xi)=0$ for both phases and every size,
introduce the independent mean and half-difference processes
$G_+=(G_{\mathrm I}+G_{\mathrm{II}})/2$ and
$G_-=(G_{\mathrm{II}}-G_{\mathrm I})/2$, and define
\begin{align}
q_+(s,s')&=A_S^2Q_S(s,s')+\tfrac12A_D^2Q_D(s,s'),
&q_-(s,s')&=\tfrac12A_D^2Q_D(s,s'),\\
d_\pm(r,s)&=\frac{r}{\ell_r^2}e^{-r^2/(2\ell_r^2)}q_\pm(s,1),
&v_\pm&=\frac{q_\pm(1,1)}{\ell_r^2}.
\end{align}
Gaussian conditioning~\cite{Rasmussen2006} gives the phase covariance
\begin{equation}
K_{\alpha\alpha'}(\mathbf x,\mathbf x')
=k_S+\delta_{\alpha\alpha'}k_D
-k_{\ell_\xi}(\xi,\xi')\left[
\frac{d_+(r,s)d_+(r',s')}{v_+}
+\eta_\alpha\eta_{\alpha'}\frac{d_-(r,s)d_-(r',s')}{v_-}\right],
\label{eq:conditioned}
\end{equation}
where $\eta_{\mathrm I}=-1$, $\eta_{\mathrm{II}}=1$, and the arguments of
$k_S,k_D$ are $(\mathbf x,\mathbf x')$. The means remain zero.
This fixes the quantum zero-temperature excess entropy while leaving the
zero-point free energies uncertain. Applying the constraint at finite size
assumes no residual zero-temperature excess entropy from cell fluctuations
after reference subtraction; the shared cell is not path-integrated.
For the time-step study, replace $K$ by
$K^{(\Delta t)}=K[1+(\kappa\Delta t)^2(\kappa\Delta t')^2]$, where
$\kappa\ge0$ has units of inverse time. The principal fit uses $\kappa=0$.

\paragraph{Free-energy predictions and active learning.}
In these coordinates,
$\mathcal L=1-rs^2\partial_r-s(1-s^2)\partial_s$.
With independent observation errors $\sigma_\ell=1.5\,\mathrm{SE}_\ell$
(defined in Section~\ref{sec:protocol}),
the observation covariance and the target--observation covariance are
\begin{equation}
\Gamma_{\ell m}=\mathcal L_{\mathbf x_\ell}\mathcal L_{\mathbf x_m}
K_{\alpha_\ell\alpha_m}(\mathbf x_\ell,\mathbf x_m)
+\sigma_\ell^2\delta_{\ell m},\qquad
[\mathbf c_\alpha(\mathbf x)]_\ell
=\mathcal L_{\mathbf x_\ell}K_{\alpha\alpha_\ell}(\mathbf x,\mathbf x_\ell).
\end{equation}
For the observation vector $\mathbf y$, the posterior mean and covariance are
\begin{equation}
\mu_\alpha(\mathbf x)=\mathbf c_\alpha(\mathbf x)^{\mathsf T}
\boldsymbol\Gamma^{-1}\mathbf y,\qquad
K^{\mathrm{post}}_{\alpha\alpha'}=K_{\alpha\alpha'}
-\mathbf c_\alpha(\mathbf x)^{\mathsf T}\boldsymbol\Gamma^{-1}
\mathbf c_{\alpha'}(\mathbf x').
\end{equation}
Evaluate at $\mathbf x_T=(k_BT/E_*,1,0)$ and add the bulk quantum harmonic
references to obtain the full phase free energies. For
$\Delta G^{\mathrm{full}}=G_{\mathrm{II}}^{\mathrm{full}}-G_{\mathrm I}^{\mathrm{full}}$,
the variance is $K^{\mathrm{post}}_{\mathrm{II,II}}+
K^{\mathrm{post}}_{\mathrm{I,I}}-2K^{\mathrm{post}}_{\mathrm{I,II}}$
at $(\mathbf x_T,\mathbf x_T)$; the cross-phase covariance is retained.

Active learning targets $Q=\Delta G^{\mathrm{full}}(T_*)$, with $T_*$ maximizing
its posterior variance on a grid covering the temperature interval of
interest and the candidate temperatures. For candidate
enthalpy $h_j=(\mathcal LG_{\alpha_j})(\mathbf x_j)$, define
$c_j=\operatorname{Cov}(Q,h_j\mid\mathbf y)$ and
$v_j=\operatorname{Var}(h_j\mid\mathbf y)$ from $K^{\mathrm{post}}$.
The predicted variance reduction and acquisition score are
\begin{equation}
\Delta V_j=\frac{c_j^2}{v_j+\widehat\sigma_j^2},\qquad
\mathcal A_j=\frac{b_j\Delta V_j}{N_jP_j(n_j+n_{j,\mathrm{equil}})},\qquad
\widehat\sigma_j^2=\frac{aT_j^2}{N_jn_j}.
\end{equation}
Here $n_j$ and $n_{j,\mathrm{equil}}$ count additional production and
equilibration steps summed over trajectories;
$a=\operatorname{median}_\ell(\sigma_\ell^2N_\ell n_\ell/T_\ell^2)$ uses
accumulated production steps. The factor $b_j=1.5$ for existing conditions
and $1$ for new ones favors extensions. Hyperparameters remain fixed
during selection and are refitted after incorporating new data.
Extensions are treated as independent precision increments for selection;
the updated mean and standard error replace the old observation on refitting.
In the benzene selection calculations, internal energies were used in place
of enthalpies, omitting the small 1-bar pressure--volume contribution; the
final free-energy fits use enthalpies.

\section{Computational details and validation}
\label{sec:protocol}

\paragraph{MTP training database.}
We generated the training database using a level-16 moment tensor
potential (MTP) and the active-learning algorithm of Novikov
et al.~\cite{Novikov2021}, with selection and stopping thresholds
$\gamma_{\rm select}=2$ and $\gamma_{\rm break}=10$, respectively.
Starting from the DFT-relaxed 48-atom structures I and II, we generated
ten perturbed configurations of each. Every Cartesian component of the
cell vectors received an independent uniform displacement in
$(-0.05,0.05)$ \AA, and each atom received an independent Gaussian
displacement with $\sigma=0.03$ \AA. Together with the two unperturbed
structures, these gave 22 initial configurations.
We then applied active learning during MTP-driven relaxations of the
perturbed structures and classical NPT molecular dynamics at 1 bar and
100 and 200 K. The MD simulations used 48-atom cells for both phases and
additional 192-atom cells for phase II. Sampling and refitting continued
until no additional configurations were selected, yielding a database
of 650 configurations.

\paragraph{Sampling and observation uncertainties.}
Table~\ref{tab:settings} summarizes the final 1-bar calculations.
The cells and bead counts form an irregular selection chosen by GP active
learning, rather than a complete grid. Each trajectory starts from its
relaxed reference, with identical nuclear positions in all beads and
distinct random seeds across the dataset.

\begin{table}[htbp]
\centering\small
\caption{Computational settings and retained data at 1 bar.}
\label{tab:settings}
\begin{tabularx}{\linewidth}{@{}lX@{}}
\toprule
Quantity & Setting \\
\midrule
Potential & Smooth level-12 MTP; 6.5 \AA\ cutoff; 20 radial functions;
271 parameters fitted to 650 DFT configurations \\
Sampling & i-PI; flexible-cell NPT; PILE-G thermostat; OBABO splitting \\
Time constants & Thermostat 100 fs; barostat 1,000 fs \\
Time step and output & 0.5 fs; energies recorded every 100 steps \\
Equilibration & First 50 ps discarded from each trajectory \\
Phase I & 38 conditions, 400 trajectories; 5--200 K;
$P=1,2,4,8$; 4, 32, 108 molecules; durations 100--2,800 ps \\
Phase II & 31 conditions, 320 trajectories; 1.25--200 K;
$P=1,2,4,8$; 16, 128, 432 molecules; durations 100--800 ps \\
Replication & Ten or twenty retained trajectories per condition \\
\bottomrule
\end{tabularx}
\end{table}

We screen final configurations to exclude trajectories showing signs of a
phase change. After periodic matching of molecular centres and removal of
overall translation, their root-mean-square displacement from the reference
must be below 0.5 \AA. For phase II, the mean ring-plane tilt must also be
below $12.5^\circ$. These are geometric endpoint criteria. Phase II appears
close to melting at 200 K; no trajectories pass for its 16-molecule cell
at $P=2,4$, so those conditions are excluded.

Let $h_{\ell a}$ be the retained time-averaged enthalpy per molecule for
trajectory $a$ at condition $\ell$, and let $n_{\mathrm{traj},\ell}$ be
the number of trajectories. Their lengths are equal within each condition.
We use
\begin{equation}
\widehat h_\ell=\frac{\sum_a h_{\ell a}}{n_{\mathrm{traj},\ell}},\qquad
\mathrm{SE}_\ell^2=\frac{\sum_a(h_{\ell a}-\widehat h_\ell)^2}
{n_{\mathrm{traj},\ell}(n_{\mathrm{traj},\ell}-1)},\qquad
\sigma_\ell=1.5\,\mathrm{SE}_\ell.
\end{equation}
Energy and volume are averaged together, preserving their covariance.
The deterministic corrections in Eq.~\eqref{eq:observation} shift the
means without changing these errors. The factor 1.5 allows approximately
for uncertainty in the estimated standard errors. Gaussian 95\% intervals
below denote $\pm2\sigma$.

\Needspace{5\baselineskip}
\paragraph{Harmonic references and fitting settings.}
At 1 bar, both the static enthalpy and vibrational spectrum use the relaxed
reference cell. Force constants are calculated by central differences
with displacements of $10^{-4}$ \AA. Displacing atoms in one 48-atom cell
and using lattice translations constructs the supercell force-constant
matrix; Fourier transformation separates it into wave-vector blocks.
All nontranslational squared frequencies are positive. Each observation
uses the spectrum for its own supercell in Eqs.~\eqref{eq:harmonic-enthalpy}
and~\eqref{eq:time-step}. Bulk quantum harmonic free energies are obtained
by linear fits in $1/N$ at each temperature, using
$N=48R^3$, $R=1,\ldots,6$, for phase I and
$N=192R^3$, $R=1,\ldots,4$, for phase II.

The nine GP hyperparameters are fitted in logarithmic coordinates by
minimizing
\begin{equation}
\frac12\mathbf y^{\mathsf T}\boldsymbol\Gamma^{-1}\mathbf y
+\frac12\log\det\boldsymbol\Gamma
+\frac12[\log(\ell_\xi/1.5)]^2,
\end{equation}
up to a constant. The last term is the prior
$\log\ell_\xi\sim\mathcal N(\log1.5,1)$, introduced to stabilize size
extrapolation during early active learning. With the harmonic scale
$\hbar\bar\omega_{\mathrm h}=0.1494$ eV (the mean mode quantum averaged
over the two largest reference cells), the search bounds are
\begin{gather}
0.02\hbar\bar\omega_{\mathrm h}\le E_*\le40\hbar\bar\omega_{\mathrm h},
\qquad 10^{-3}\hbar\bar\omega_{\mathrm h}\le E_*\ell_r
\le10\hbar\bar\omega_{\mathrm h},\\
10^{-3}\le\ell_{s,S},\ell_{s,D}\le10,\qquad
10^{-4}\le\ell_\xi\le10^3,\qquad
10^{-5}\le A_S,A_D,B_S,B_D\le1.2\times10^5.
\end{gather}
The amplitude bounds are in meV per molecule.
Multiple starts, including previous fits, are used; the best objective is
reproduced from distinct starts. The fitted $E_*$ is 14.975 meV.
Reported GP uncertainties are conditional on the fitted hyperparameters
and harmonic references; they exclude errors in the potential and harmonic
size extrapolation.

\paragraph{Student-$t$ sensitivity.}
We replace the Gaussian sampling-error model by
$(y_\ell-f_\ell)/\mathrm{SE}_\ell\sim t_{\nu_\ell}$, where
$f_\ell=(\mathcal LG_{\alpha_\ell})_\ell$ and
$\nu_\ell=n_{\mathrm{traj},\ell}-1$ is 9 or 19. The scale is
$\mathrm{SE}_\ell$, without the Gaussian model's factor of 1.5.
GP hyperparameters remain fixed at their Gaussian fit values. Using the
Gaussian scale-mixture representation of the Student-$t$
likelihood~\cite{Jylanki2011}, we sample latent observation precisions and
enthalpies with two independent Gibbs chains of 20,000 iterations each,
discard the first 2,000, and retain every tenth state. The 3,600 conditional
Gaussian predictions are equally weighted; agreement between the chains'
free-energy quantiles checks convergence. The median phase difference
changes little, while 95\% intervals narrow by 12\% on average over
0--200 K. This is consistent with a $t_9$ standard deviation of
$\sqrt{9/7}\,\mathrm{SE}\simeq1.13\,\mathrm{SE}$, below the Gaussian
model's $1.5\,\mathrm{SE}$.

\begin{figure}[H]
\centering
\includegraphics[width=1\linewidth]{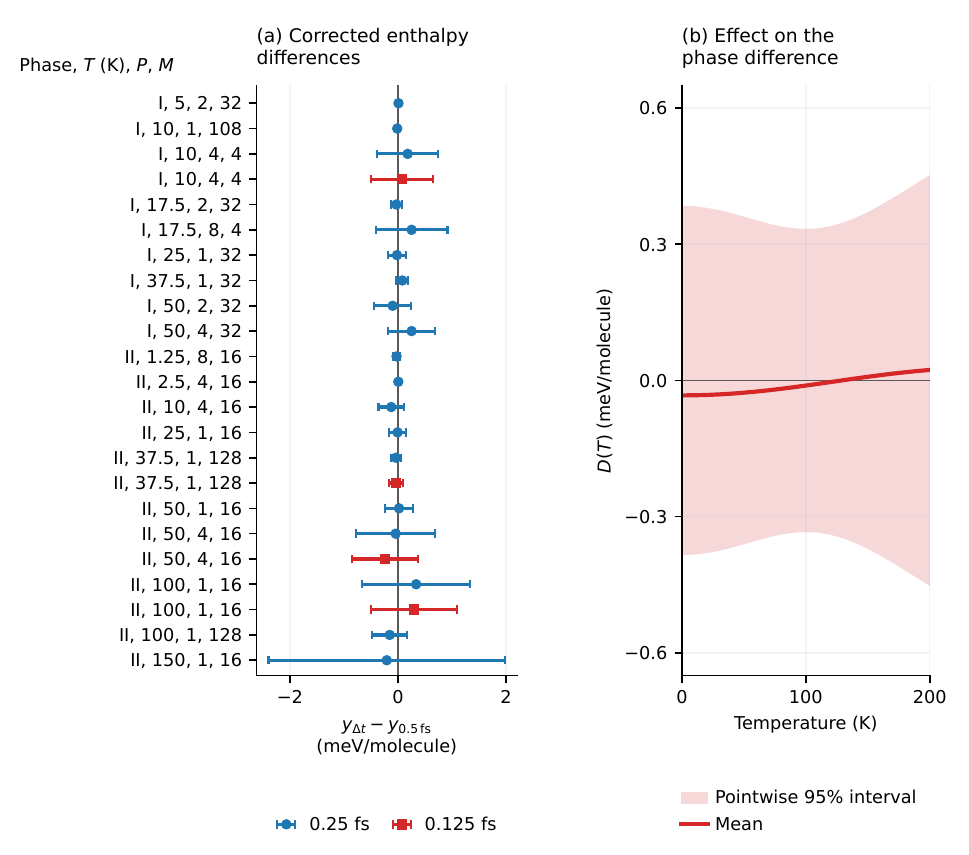}
\caption{Time-step convergence after harmonic bias correction.
(a) Excess enthalpy at 0.25 or 0.125 fs minus its value at 0.5 fs.
Labels identify phase, temperature $T$, bead count $P$, and molecule count $M$.
Bars show twice the standard deviation of the difference, adding the
two observation variances with $\sigma_\ell=1.5\,\mathrm{SE}_\ell$ and
treating sampling errors as independent.
(b) Change $D(T)$ in the quantum bulk phase free-energy difference upon
extrapolating to zero time step, using all 92 observations. The line is
the mean after averaging over $\kappa$; at each temperature the shaded
interval is symmetric about zero and contains 95\% of the posterior
probability for $D(T)$.}
\label{fig:timestep}
\end{figure}

\paragraph{Time-step convergence.}
We continue selected trajectories at 0.25 and 0.125 fs, discard 10 ps after
the change, and average a further 50--400 ps in ten trajectories per
condition. The original runs last at least 100 ps; conditions still showing
slow structural relaxation are excluded. The 23 additional observations
cover both phases, 1.25--150 K, $P=1,2,4,8$, and 4--128 molecules.
Figure~\ref{fig:timestep}a compares their corrected enthalpies.

We fit the combined 92 observations with $K^{(\Delta t)}$ and a uniform
prior on $\kappa\ge0$. At each $\kappa$, the other hyperparameters are
refitted, and the conditional free-energy distributions are averaged
with weights proportional to the fitted marginal likelihood.
Define the time-step effect using predictions from this same dataset:
\begin{equation}
D(T)=\Delta G_{\mathrm q}^{\mathrm{full}}(T;0)
-\Delta G_{\mathrm q}^{\mathrm{full}}(T;0.5\,\mathrm{fs}),
\end{equation}
where the subscript q denotes $P,N\to\infty$.
We obtain $D(0)=-0.03\pm0.17$ meV per molecule (one posterior standard
deviation). Averaging over $\kappa$ increases the standard deviation of
$\Delta G_{\mathrm q}^{\mathrm{full}}(T;0)$ by at most 11\% relative to fixing
$\kappa$ at its best fit. No residual time-step dependence is detected
at this precision. The augmented dataset gives a zero-temperature gap
of $17.57\pm0.53$ meV per molecule, compared with $18.18\pm0.60$ from
the principal 69 observations. The shift mainly reflects additional
sampling; the principal results use the 69-observation dataset.

\bibliography{references}